\documentclass[pdflatex,sn-aps]{sn-jnl}

\usepackage{braket}
\usepackage{array}
\usepackage{graphicx}%
\usepackage{multirow}%
\usepackage{amsmath,amssymb,amsfonts}%
\usepackage{amsthm}%
\usepackage{mathrsfs}%
\usepackage[title]{appendix}%
\usepackage{xcolor}%
\usepackage{textcomp}%
\usepackage{manyfoot}%
\usepackage{booktabs}%
\usepackage{algorithm}%
\usepackage{algorithmicx}%
\usepackage{algpseudocode}%
\usepackage{listings}%

\theoremstyle{thmstyleone}%
\theoremstyle{thmstyletwo}%

\theoremstyle{thmstylethree}%

\begin{document}

\title[Article Title]{Towards Nonlinear Quantum Thermodynamics}


\author[1]{\fnm{Gershon} \sur{Kurizki}}\email{gershon.kurizki@weizmann.ac.il}

\author*[2]{\fnm{Nilakantha} \sur{Meher}}\email{nilakantha.meher6@gmail.com}

\author[3]{\fnm{Avijit} \sur{Misra}}\email{avijitmisra@iitism.ac.in}

\author[4]{\fnm{Durga Bhaktavatsala} \sur{Rao Dasari}}\email{d.dasari@pi3.uni-stuttgart.de}

\author[5]{\fnm{Tomas} \sur{Opatrny}}\email{tomas.opatrny@upol.cz}

\affil[1]{Department of Chemical and Biological Physics \& AMOS,Weizmann Institute of Science, Rehovot 7610001, Israel}

\affil*[2]{Department of Physics, SRM University AP, Amaravati 522240, Andhra Pradesh, India}

\affil[3]{Department of Physics, Indian Institute of Technology (ISM), Dhanbad, Jharkhand 826004, India}

\affil[4]{3. Physikalisches Institut, Center for Applied Quantum Technologies, University of Stuttgart, Stuttgart, 70569, Germany}

\affil[5]{Palacky University, 77146 Olomouc, Czech Republic}


\abstract{We have recently put forth several schemes of unconventional, nonlinearly-enabled thermodynamic (TD) devices that can operate in either the classical or the quantum domain by transforming  thermal-state input in multiple uncorrelated modes into non-gaussian state output in selected modes: a four-mode Kerr-nonlinear interferometer that acts as a heat engine; two coupled Kerr-nonlinear  Mach-Zehnder interferometers that act as  a phase microscope  with  unprecedented phase resolution; and a noise  sensor that can distinguish between unknown nonlinear quantum processes.  These schemes reveal the unique merits of nonlinear TD devices: their ability to act in an autonomous, fully coherent,  dissipationless fashion, unlike their conventional counterparts. 
Here we present the opportunities and challenges  facing this new paradigm of nonlinear (NL)  quantum  and classical TD devices along the following lines: 
A) Linear versus nonlinear multimode transformations in TD devices:  what are the principal distinctions between the two types of transformations? B) Classical versus quantum effects in NL TD devices: what are their main differences? Is quantumness an advantage or a disadvantage? C) Deterministic methods of achieving giant nonlinearity at the few-photon level  via coherent processes, including multiatom-bath  interactions which can paradoxically yield NL Hamiltonian effects: their comparison with probabilistic, measurement-based methods that can achieve similar NL effects in the quantum domain.}





\maketitle
\section{Introduction}
\label{Sec1}
The ongoing extension of thermodynamics to the quantum domain \cite{kurizki2022Book,BinderBook} has been particularly prominent in the introduction of quantum schemes for thermodynamic (TD) devices such as engines and refrigerators \cite{kurizki2022Book,BinderBook,scovil1959three,Scully2003Science,Scully2011PNAS,Rossnagel2014PRL,niedenzu2018quantum,cangemi2024quantum,Ghosh2017,Ghosh2018,ferreri2024quantum,guff2019power,scully2010quantum,
hammam2022exploiting,klaers2017squeezed,niedenzu2015performance,niedenzu2018cooperative,rolandi2023collective,jaramillo2016quantum,
hartmann2020many,zhang2022work,fogarty2020many,souza2022collective,b2020universal,rossnagel2016single,kosloff2014quantum,uzdin2015equivalence,kosloff1984quantum,Allahverdyan2004EPL,ARQST,Gardas2015PRE,CarnotPRE,mukherjee2020anti,xu2022minimal}, diodes and transistors \cite{PhysRevE.99.042121,Tahir,segal2005spin,saira2007heat,segal2008single,shen2011single,meher2019atomic,
karimi2017coupled,ronzani2018tunable,joulain2016quantum}. Yet, despite the fact that many of these schemes involve quantum coherence and squeezing \cite{Ghosh2017,Ghosh2018,ferreri2024quantum,guff2019power,scully2010quantum,
hammam2022exploiting,klaers2017squeezed,uzdin2015equivalence,AREPL,Mukhopadhyay2018PRE} or quantum cooperativity \cite{niedenzu2015performance,niedenzu2018cooperative,rolandi2023collective,jaramillo2016quantum,
hartmann2020many,zhang2022work,fogarty2020many,souza2022collective,b2020universal}, they are all based on {\it dissipative open systems }\cite{kurizki2022Book,BinderBook,breuer2002theory}. Hence, their TD description disqualifies them from acting as fully quantum mechanical devices that are rules by unitarity.

Aiming to break away from this established paradigm of TD devices as dissipative open systems, we have recently introduced the concept of nonlinear (NL) thermodynamics in both quantum and classical domains. It is concerned with NL effects on field modes that are fed with thermal  noise at the input to TD engines or sensors \cite{Opatrny2023ScAdv,meher2024thermodynamic,meher2024supersensitive,kurizki2025nonlinearity}.  Since each mode is then in a maximal- entropy, passive state, these input modes are neither work nor information resources \cite{pusz1978passive,Allahverdyan2004EPL,Gardas2015PRE, PRE2014}. Yet if the modes are coherently transformed/mixed  or filtered by NL elements within the device,  some of them may become non-thermal  and  constitute  useful TD resources for work production in   selected output modes  that attain non-passive, non-gaussian states, or for quantum-enhanced sensing, by virtue of the high information these states carry \cite{Opatrny2023ScAdv,meher2024thermodynamic,meher2024supersensitive,kurizki2025nonlinearity}. 
  Such NL- filtering  thermodynamic  (TD) devices are treated here as {\it non-dissipative and fully coherent} to a good approximation. They obey the second law of thermodynamics \cite{CarnotBook,clausius1879mechanical,kurizki2022Book,BinderBook}, which excludes entropy reduction in the overall multimode output as compared to the total input,  yet  allows energy and entropy reshuffling among the modes, which is essential for the device functionality.  Such redistribution is accomplished in these NL devices {\it autonomously}, without external intervention or control.  Importantly, since heat baths are replaced in the conceived NL coherent devices by single-mode fields, the appropriate thermodynamic quantities can be inferred from the output photon statistics (compare with \cite{han2024quantum,han2025prospect}).
  
Here we briefly survey  the principles of NL- filtering, coherent TD devices, focusing on  multimode interferometers  with NL elements  that can act as heat engines or quantum sensors (Sec. \ref{Sec2}). We then compare the principles of such devices at the quantum and classical level, elucidating the limitations and advantages of their quantumness (Sec. \ref{Sec3}). This article is, however, mostly dedicated to possible methods of achieving the giant nonlinearity required for NL filtering at the few-photon level. These methods  are  allowed by the present technological state-of-the-art to be fully unitary/coherent, hence deterministic (Sec. \ref{Sec4}). Yet, as a substitute for these technically demanding  methods, one may use probabilistic, measurement-based methods that are shown to provide a viable alternative (Sec. \ref{Sec5}). The discussion (Sec. \ref{Sec6}) summarizes the  results and  presents  an outlook to further developments of the NL-filtering  thermodynamic approach.

\section{Principles and implementations of  NL filtering TD devices}
\label{Sec2}
A NL coherent TD device, be it a heat engine or a sensor, is  governed in the quantum description by a product of unitary evolution operators that have the schematic  form 
\begin{subequations}\label{Unitary}
\begin{align}
\hat U=\hat U_{L_{out}}\hat U_{NL}\hat U_{L_{in}}
\end{align}
The overall  transformation of the input state $\rho_{in}$ to the  output state is then 
\begin{align}
    \rho_{out}=\hat U \rho_{in} \hat U^\dagger
\end{align}
\end{subequations}
Here $\hat U_{L_{in}}$ and $\hat U_{L_{out}}$  stand for  the  linear transformations of the input block  into the output block   via  coherent  multimode mixing, and  $\hat U_{NL}$ is  the  multi-mode  transformation that is responsible for the NL filtering in the device.

The  linear transformation operators, effected by conventional beam splitters and phase shifters,  constitute rotations on the Poincare sphere of  mode-pair states or distributions \cite{Gerry,Hofman2009PRA}. Such linear transformations   preserve the gaussian character of the individual-mode states. By contrast,  a NL transformation, particularly the  two-mode  cross-Kerr (CK) transformation 

\begin{align}\label{CKUnitary}
  \hat U_{CK}= e^{i\chi \hat a^\dagger \hat a \hat b^\dagger \hat b}   
\end{align}
that couples the photon-number operators of modes $a, b$ with strength $\chi$, can generally twist and distort  the mode-pair phase-space distributions and observables \cite{Kitagawa1993PRA,Opatrny2015PRA,Opatrny2015PRA2}. 

Such NL transformations  can  render the state of a chosen mode non-gaussian and  quantum-correlated/entangled with other modes. 
The input state of each mode may be taken to be thermal, but it is essential that not all input states be at the same temperature,  to  ensure the non-passive character of the overall multimode input  state: the simplest example that adheres to this condition is an  input  comprised of empty (zero-temperature) modes along with hot modes. Under this condition,  a unitary  NL transformation as in Eqs. \eqref{Unitary}, \eqref{CKUnitary}  does not create ergotropy \cite{kurizki2022Book,Allahverdyan2004EPL}, but rather redistributes the entropy and energy among the modes in  a manner  compatible with the second law,  such that the  states of chosen modes become non-passive, i.e. acquire ergotropy  and hence are capable of work production \cite{pusz1978passive,Allahverdyan2004EPL} whereas other modes heat up. 
In what follows, we briefly illustrate the manifestation of  these principles  in  three generic types of NL filtering TD devices: heat engines , phase estimators and quantum noise sensors. 

\subsection{NL- filtering heat engine:}    
\begin{figure}
\includegraphics[scale=0.28]{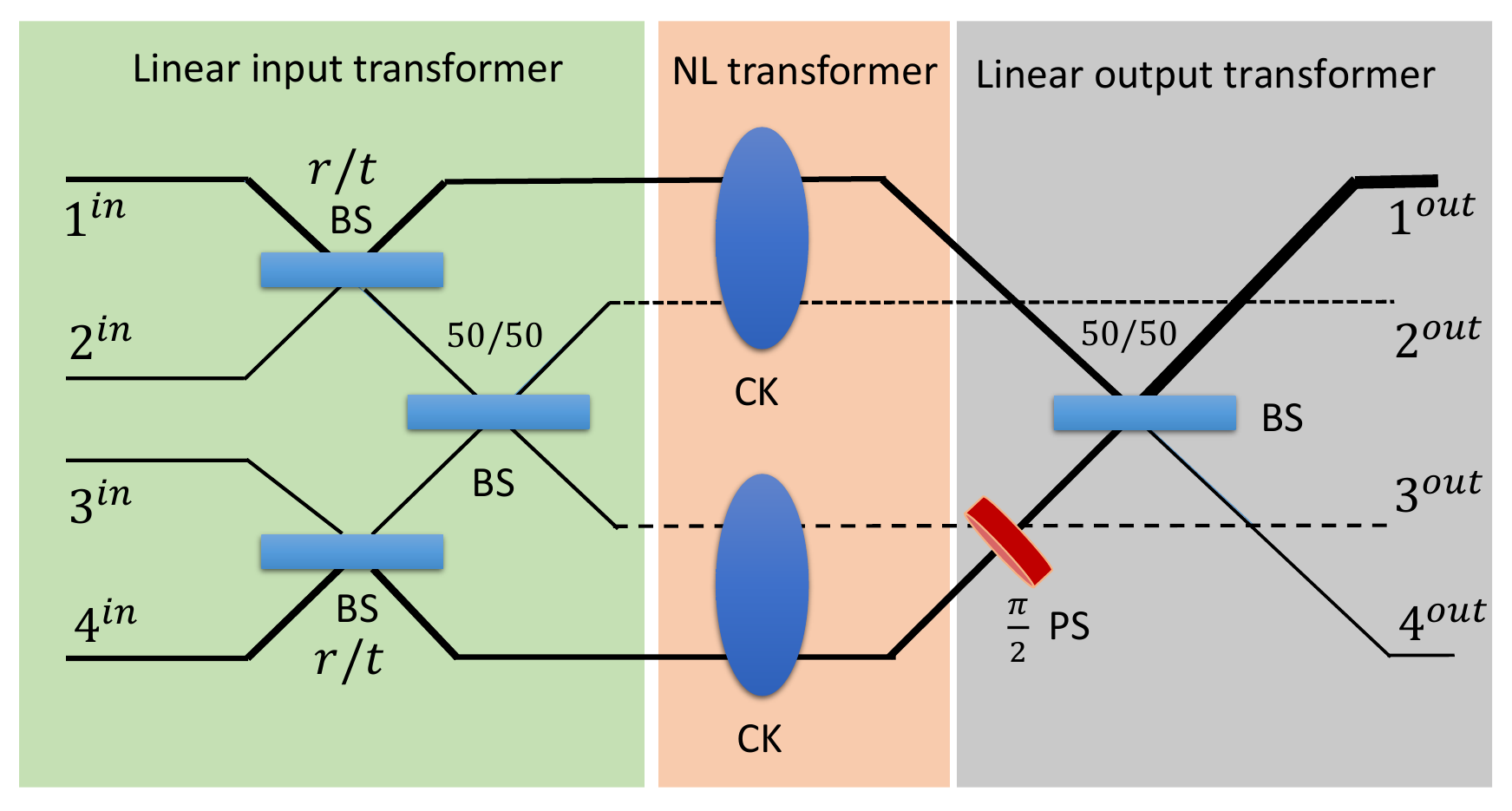}
\caption{Nonlinear coherent heat engine. The input contains two hot modes (modes $1^{in}$ and $4^{in}$) and two cold modes (modes $2^{in}$ and $3^{in}$ ) mixed by linear input transformer. Upon choosing the parameters appropriately, energy is concentrated in mode $1^{out}$ due to constructive interference between the correlated modes in the NL transformer, followed by steering in a linear output transformer.}\label{NCHE} 
\end{figure}

As an example of such an engine, our group analyzed \cite{Opatrny2023ScAdv} a four-mode interferometer fed by two high-temperature and two zero-temperature (empty) input modes (Fig. \ref{NCHE}), modes having the same frequency degenerate. Whereas standard  interferometers act as linear transformer that cannot change the output ratio of two modes with equal temperatures, one may control  their output ratio upon  adding  two NL cross-Kerr(CK) elements that create classical correlations or quantum entanglement of all four modes. 

One  can analyze the resulting NL interference in such a device  by viewing the input states in the hot modes 1 and 4  as thermally mixed coherent states that have randomly  distributed mean phases and a gaussian distribution of mean amplitudes. For each coherent-state component of these thermal states,  the mean intensities in the output (out) modes $1^{out}$ and $4^{out}$, $|\alpha^{out}_{ 1,4}|^2$, are obtained  from their input counterparts by the following relation:  
\begin{align}\label{meanoutputcoherent}
|\alpha_{1,4}^{out}|^2=\frac{r^2}{2}[(\alpha_1^2+\alpha_4^2)_{in}\pm (\alpha_1\alpha_4)_{in} \sin(2t^2\alpha_1\alpha_4\chi\cos\phi-\phi)_{in}],
\end{align} 
Here the CK coupling strength $\chi$ is as in Eq. \eqref{CKUnitary}, $t^2$ and $r^2$ are the respective  transmissivity and reflectivity of the input beam- splitter (BS)  (see Fig. \ref{NCHE}). The interference term in Eq. \eqref{meanoutputcoherent} is non-sinusoidal in the phase difference $\phi$ of the hot input modes. This term reduces to $\sin\phi$ when $\chi = 0$, resulting in interference washout following averaging over random $\phi$. This differs from the case of appreciable CK nonlinearity which gives rise to a narrow-peaked phase $\phi-$ distribution and, if  $\chi$ and $t$ are chosen appropriately,   causes destructive interference in output mode $4^{out}$, and constructive interference in output mode $1^{out}$, so that the mean intensity is steered  from  input mode $4^{in}$ to  output mode $1^{out}$. 
Averaging over random input phases  $\phi$  and a thermal distribution of the hot input modes, yields the following mean intensities (photon numbers)  of the hot output modes satisfy \cite{Opatrny2023ScAdv}
\begin{align}
\bar{n}_{1,4}^{out}=r^2 \bar{n} \left[1\pm \frac{t^2 \chi \bar{n}}{(1+t^4 \chi^2\bar{n}^2)^2} \right].
\end{align}
 The outcome is then energy amplification in the chosen output mode $1^{out}$ while the overall energy is conserved.   Equally  important is the transformation of the input thermal state in mode $1^{in}$  to a non-passive, non-gaussian  state  capable of delivering work at the output mode $1^{out}$  since the NL filtered ergotropy is concentrated in this mode, while entropy  increases in all other  output modes. 
 
The adherence of this device  to the second law of thermodynamics  has been verified, analytically and numerically:  the analysis shows that the combined  four-mode input state is non-passive, although each input mode is in a thermal (passive) state, hence ergotropy is  merely redistributed and the sum of entropies of the individual modes increases while the overlall entropy is conserved by unitarity (coherent) \cite{Opatrny2023ScAdv}. This NL-filtering  redistribution enables the device to operate as a heat engine capable of coherently transforming heat to work in a designated mode.  The extracted work can be measured via power detection resolved at the single-photon level, as has been demonstrated in a superconducting circuit \cite{cottet2017observing}.

\subsection{ NL-filtering thermal  quantum sensors:}  
\begin{figure}
\includegraphics[scale=0.4]{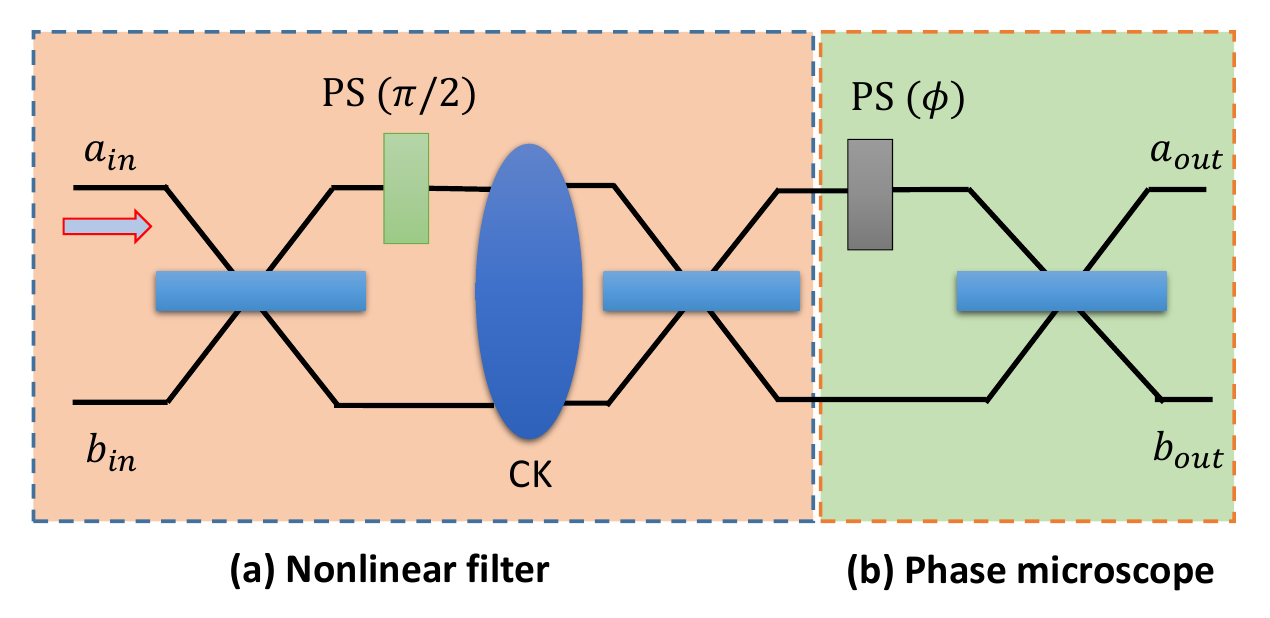}
\caption{(a) Nonlinear filter serves as input to (b) transmission microscope aimed at estimating an unknown phase shift (PS) $\phi$. }\label{PRA1} 
\end{figure}

NL mode couplings  have been  shown by us \cite{meher2024supersensitive}  to allow  supersensitive phase estimation (SSPE)  in interferometers illuminated by few photons. Such SSPE can be used in a  quantum phase sensor  that functions as a transmission microscope. It is  based on coupled Mach-Zehnder interferometers  (MZIs) in which a sample is  detected by means of the phase shift $\phi$ of lradiation transmitted  along one of the arms in the MZI (Fig. \ref{PRA1}). For classical-like light, the resolvable phase exceeds the standard quantum limit (SQL) \cite{Gerry,Carmichael_BOOK,Gardiner,
ScullyZubairy} $\Delta\phi \geq 1/\sqrt{\bar{n}}$.  The  SQL can, however, be violated  by resorting to entangled  N00N two-mode states \cite{PhysRevLett.104.123602} $(\ket{n, 0} + \ket{0, n})/\sqrt{2}$ comprised of $n$ photons in one mode and 0 in the other, yet these states are hard to implement and easy to destroy by loss and decoherence, particularly for $n \gg 1$.

We pointed out \cite{meher2024supersensitive} that  instead of  pure-state N00N input, one may resort to thermal-noise input subject to NL filtering (Fig. \ref{PRA1}). The interferometer  phase resolution is then much below the SQL and even lower than the  nominal Heisenberg  uncertainty limit (HL) \cite{Gerry,Carmichael_BOOK,Gardiner,
ScullyZubairy}  $\Delta\phi \sim 1/\bar{n}$, since the HL does not hold for distributions with small $\bar{n}$ and a large variance, $\Delta n > \bar{n}$, which  may allow for   $\Delta\phi < 1/\bar{n}$ \cite{Monras2006PRA,Hofman2009PRA,Pinel2013PRA,Boixo2007PRL,Giovannetti2012PRL}.

The   phase resolution  is limited by the quantum Cramer-Rao bound \cite{cramer1999mathematical}, whereby  the minimal phase error that can be reached by a given input state  satisfies

\begin{align}\label{CramerRao}
\Delta\phi_{min} \geq \frac{1}{\sqrt{F_Q}}.
\end{align}
$F_Q$ being  the quantum Fisher information (QFI)  that maximizes the phase-shift estimation accuracy \cite{birrittella2021parity}.  FQ is determined by the output state of the coupled MZI as in Fig. \ref{PRA1}, whose general form is as in Eq. (\ref{Unitary}b), upon allowing for $\hat U_{NL}$ and the phase shift (PS) operation $\hat U_{PS}(\phi) = e^{i\phi \hat a^\dagger \hat a}$ that  corresponds to   the unknown phase shift $\phi$ in mode $a$.  In a coupled MZI with cross-Kerr (CK) nonlinearity, $\hat U_{NL}=\hat U_{CK}$, as in Eq. \eqref{CKUnitary}. The  CK  evolution operator is responsible for the NL filtering in this device. 

\begin{figure}
\includegraphics[scale=0.4]{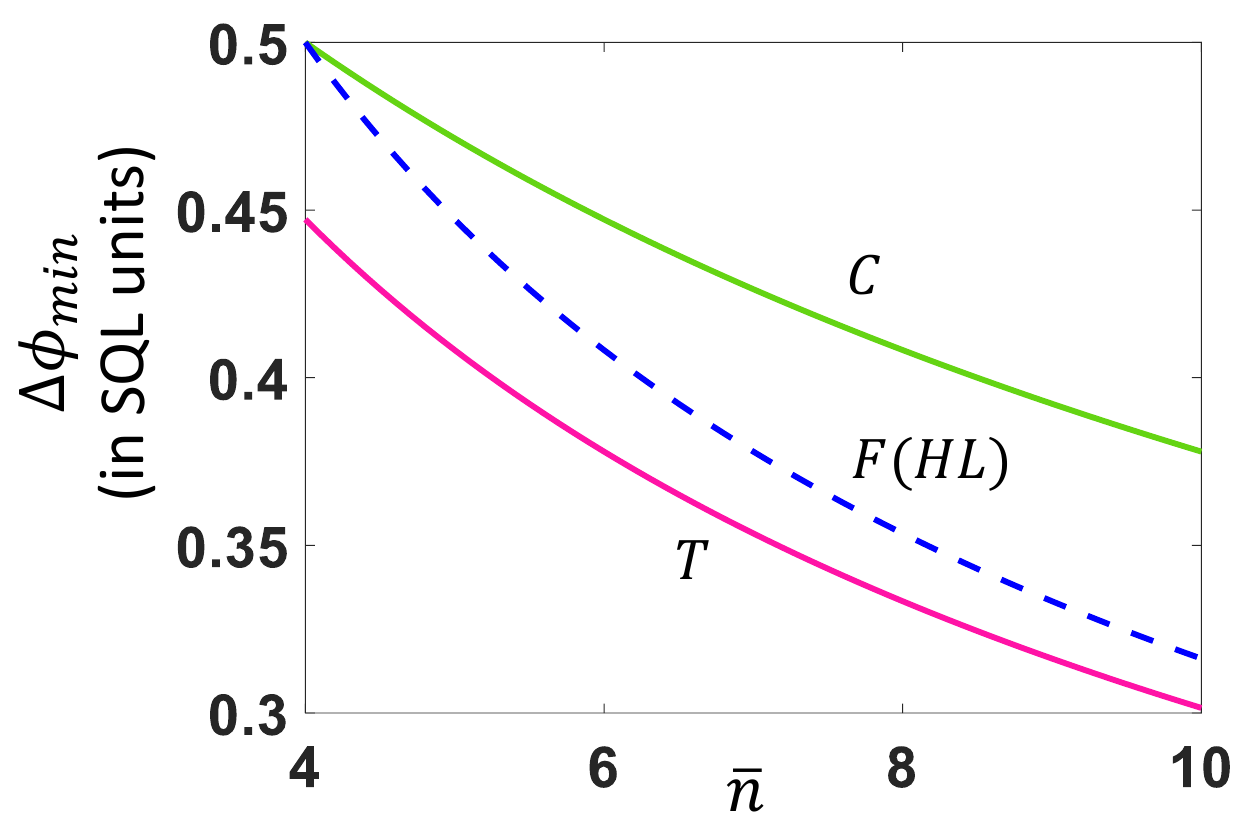}
\caption{ The minimal
phase error normalized to the SQL error as a function of the mean input photon number for thermal (T), coherent (C) and Fock-state (F) inputs to the setup in Fig. \ref{PRA1}. Here the CK NL phase shift has the optimal value per photon $\chi=\pi/2$. The F input yields the HL phase error. }\label{PRA2} 
\end{figure}

We have obtained  solutions for $F_Q$ corresponding to  thermal (T), coherent (C) and Fock-state (F) inputs  that are NL- filtered by this coupled MZI, provided the mean input  photon number $\bar n$ exceeds 4 (Fig. \ref{PRA2})  
\begin{align}
(F_Q)_{T}= \bar{n}^2+\bar{n}  > (F_Q)_{F}=\bar{n}^2 > (F_Q)_{C}=\tfrac{1}{2}\bar{n}^2+2\bar{n}.
\end{align}

This surprising result shows that thermal-noise input filtered by the NL(CK) transformation allows higher phase sensitivity than that obtainable through the same NL filtering using Fock- or coherent-state input having equal mean photon numbers [Fig. \ref{PRA2}]. This surprising advantage of thermal input has the following reason: the broad Fock state distribution of thermal noise is NL-transformed  to a wider distribution of N00N states than its Fock- or coherent-state counterparts, and since the phase sensitivity bound is determined from the quantum Fisher information (QFI) by the highest N00N state, the broadest spread yields the phase resolution. 

This finding opens a new path to SSPE using thermal light sources. Remarkably,   this SSPE persists even in the presence of high losses in the coupled MZI setup \cite{meher2024supersensitive}.

 \subsection{NL quantum noise sensors:}  NL MZIs endow thermal input in certain modes with information content   that  may be employed  to characterize  the Hamiltonian that governs the \textquotedblleft black box\textquotedblright medium \cite{Chuang1997JModOpt} inside  the MZI. This unknown Hamiltonian \cite{Goldman2014PRX,del2023dynamical} gives rise to $k$-photon interaction between the MZI modes, and have, e.g, the form 
\begin{align}\label{kphoton}
\hat U_{NL}^{(k)}=e^{-igt(\hat a^{\dagger k} \hat b^k+\hat a^k \hat b^{\dagger k})},
\end{align}
with unknown  two-mode coupling strength $g$, as well as unknown NL-order  $k$.

We have shown \cite{meher2024thermodynamic} that  instead of time-consuming tomography \cite{Chuang1997JModOpt,Goldman2014PRX,del2023dynamical,Mohseni2008PRA,Lvovsky2009RMP} as a means of characterizing   this   quantum NL interaction,  it  may be inferred from the  time-dependent ergotropy/ work capacity of an  output mode that  has been transformed from a gaussian, passive input state to a non-passive, non-gaussian, output state. Importantly, ergotropy does not arise if the underlying noise in the MZI is caused by linear or Raman two-mode coupling, given that the input in each mode  is in a  thermal, or, more  generally,  passive, state \cite{pusz1978passive,Allahverdyan2004EPL,uzdin2018global,kurizki2022Book}.

Our proposed  NL quantum noise sensor illustrates the vast  potential of NL filtering: it allows, by  single-mode ergotropy probing, to fully  characterize unknown  two-mode  quantum noise properties.

\section{ Quantum versus   classical NL filtering} 
\label{Sec3}
The foregoing survey attests to the possible advantageous effects and applications  of NL-filtered thermal noise whose  quantum and classical  principles  have been briefly outlined above. Here we ask: which of the NL filtering effects are exclusively quantum? By contrast, which effects can be both quantum and classical? What are the advantages or disadvantages of operating in the  quantum domain?

 We first address these questions in the context of  our 4-mode  heat engine (Sec. \ref{Sec2}).
Quantum correlations of  the  NL-filtered 4-mode fields  are rendered by the Stokes operators \cite{varshalovich1988quantum,leonhardt1997measuring} which  can be expressed in terms of the creation and annihilation operators  of mode -pair  $i$ and $j$, as $ \mathcal{\hat J}_+^{(ij)}=\hat a_i^\dagger \hat a_j$, $\mathcal{\hat J}_-^{(ij)}=\hat a_i \hat a_j^\dagger$, $\mathcal{\hat J}_z^{(ij)}=\tfrac{1}{2}(\hat a_i^\dagger \hat a_i-\hat a_j^\dagger \hat a_j)$, $\mathcal{\hat J}_0^{(ij)}=\tfrac{1}{2}(\hat a_i^\dagger \hat a_i+\hat a_j^\dagger \hat a_j)$.    Under NL transformations, these Stokes operators undergo  non-Gaussian twisting  that entangles  all  4 modes and creates complicated quantum correlated states \cite{Kitagawa1993PRA,Opatrny2015PRA,Opatrny2015PRA2}.

\begin{figure*}
\includegraphics[scale=0.4]{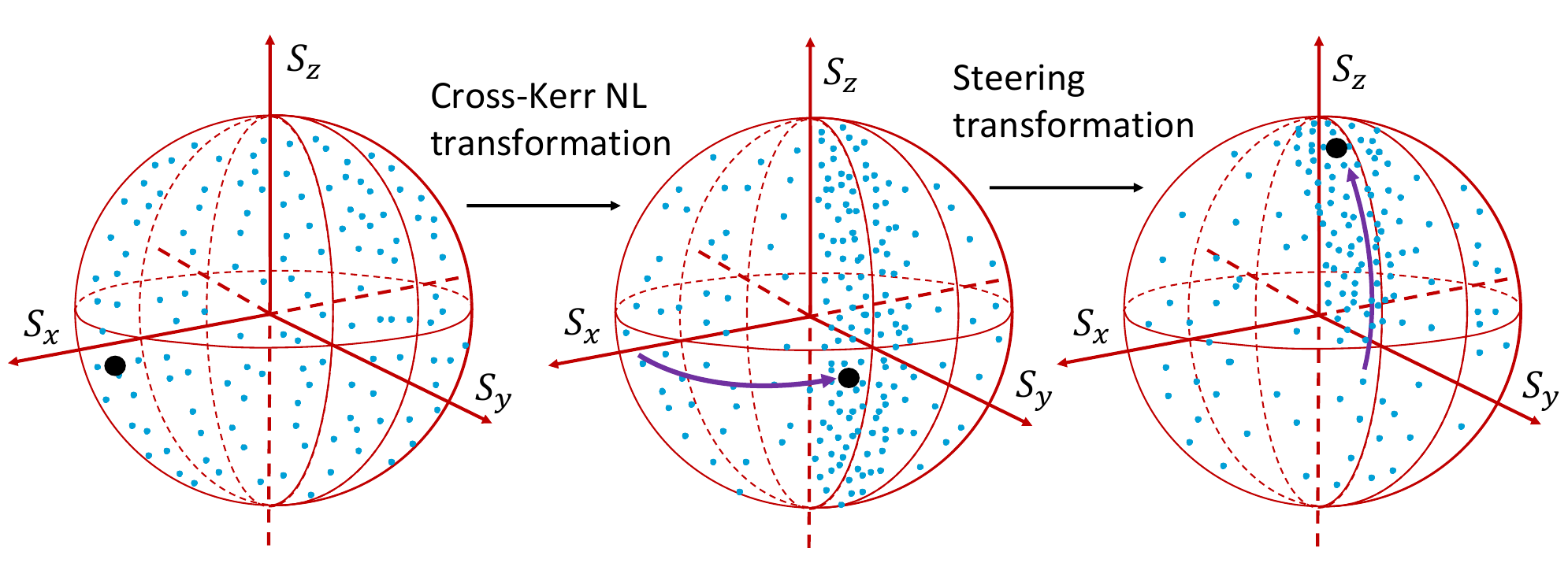}
\caption{NL transformation in phase space: Dots denote  coherent states that are randomly distributed on the Poincare sphere in the thermal ensemble. The black point denotes a randomly picked state, The violet arrows denote the dots concentration by NL CK (middle) and linear steering by BS and PS (right). }\label{ScAdv2} 
\end{figure*}

Classical  stochastic  mode correlations  are described  by the  ensemble-averaged Stokes parameters, which are the mean values of the Stokes operators described above. For the hot  output modes $1^{out}$ and $4^{out}$, $S_k=\langle \mathcal{\hat J}_k^{(14)}\rangle $, $k=x,y,z,0$.  Since  a  two-mode coherent state corresponds to  a point  on the surface of the  Poincare sphere,   their random gaussian distribution in a thermal-ensemble is merely rotated on the sphere surface by  linear gaussian operations. By contrast, the NL CK  coupler  deforms and concentrates the random phase-space distribution in the $S_y < 0$ region of  the  sphere surface (Fig. \ref{ScAdv2}). This  distribution can be  subsequently steered  by  further linear operations  to the $S_z > 0$ part  of the sphere surface, which corresponds to the mean energy amplification  in a chosen mode (here mode $1^{out}$). Concurrently,  entropy is reduced   and   ergotropy,  that  is the capacity  for work production \cite{pusz1978passive,Allahverdyan2004EPL,uzdin2018global,kurizki2022Book}, is increased in that mode. These redistributions satisfy the first and second laws of thermodynamics in the closed 4-mode system, since the mean-energy is steered to the chosen mode (here from mode $4^{in}$ to mode $1^{out}$)   and the entropy reduction in  that mode comes at the expense of entropy growth in the  undesirable  modes (here  modes 2 and 3). 

 In order to compare the above classical and quantum descriptions, let us first  assume that  the  input modes $1^{in}$ and $4^{in}$ are populated by coherent states with equal mean amplitudes, $\alpha_1 = \alpha_4$, and random  mean phases. The mean quanta numbers in  the corresponding  output modes, upon averaging over these random phases,  are then obtained from a quantum calculation in the form \cite{Opatrny2023ScAdv}
 
\begin{align}\label{MeanPhoton14}
   \overline{ \langle \hat n_{1,4}^{out}\rangle }=\bar{S}_0^{({out})}\pm \bar{S}_z^{({out})}=r^2\alpha_1^2 (1\pm J_1(b) e^{-d}).
\end{align} 
Here  $\hat n_{i}=\hat a_i^\dagger a_i, d=t^2(1-\cos \chi)(\alpha_1^2+\alpha_4^2)$ and $J_1(b)$ is the first-order Bessel function with argument $b = 2t^2\alpha_1\alpha_4\sin \chi$. 

In the corresponding  classical  calculation \cite{Opatrny2023ScAdv}, Eq. \eqref{MeanPhoton14} is revised to have $d = 0$. 
Thus, we find a  disadvantage  of the quantum NL transformation compared to  its classical counterpart: the nonzero $d$ value, which stems  from vacuum fluctuations of the  fields in the quantum description, exponentially decreases the energy steering in Eq. \eqref{MeanPhoton14}.

On the other hand, quantum calculations have the unique advantage of allowing us to monitor the changes in the input  photon statistics caused by NL transformation effects, which underlie the applications discussed in Sec. \ref{Sec2}, particularly  the ability to improve phase estimation  \cite{meher2024supersensitive} and sense quantum noise via output-mode ergotropy \cite{meher2024thermodynamic}.

Thus, whereas  NL-filtered mean-energy flow between modes may  feature quantum disadvantage, as illustrated above, there is no substitute for quantumness in applications based on NL transformations of input quantum-statistics: these strictly quantum transformations  may endow selected modes with ergotropy and increased information content at the expense of other, unused, modes.

\section{Methods of  deterministic  NL filtering in the quantum domain}
\label{Sec4}
Progress towards NL quantum thermodynamics is contingent on the realization of NL effects at the level of a few photons. Two possible giant NL effects are discussed in what follows: 
\subsection{Dipole-dipole interactions of Rydberg polaritons:} 
A giant NL CK effect, which was previously predicted by our group \cite{Friedler},  has been lately demonstrated \cite{drori2023quantum} in a cold rubidium trap,  where each photon  is converted by a pump field  into a Rydberg atom polariton. In this setup, counter-propagating photon pairs  become cross- correlated via dipole-dipole interactions over distances of the order of the photon wavelength \cite{Friedler,drori2023quantum} . These long-range interactions gives rise to CK phase shifts of $\pi$ per photon. Thus Rydberg polaritons can be strong  enough to impose NL transformations  on  few-photon thermal states.

\subsection{NL multiatom transformations  by linear coupling to a thermal bath:} Bath-mediated, dispersive interaction among  multiple spins or atoms  has been shown by our group \cite{bhaktavatsala2011generation}  to  result in   an effectively unitary  NL  state transformation, even though  the  multispin/multiatom system-bath interaction is linear, as usual. This NL unitary evolution can yield, with high probability,  entangled, non-gaussian states, particularly macroscopic quantum superposition (MQS) states,  of $N$ atoms  or spins that are collectively coupled to a bosonic bath, even if the bath is at finite temperature.

The scenario involves $N$  two-level systems (TLS) that interact with a bosonic (photonic or phononic) bath. It  is governed by the collective many-body Hamiltonian
\begin{align}\label{AtomBathHamiltonian}
    &\hat H=\hat H_S+\hat H_B+\hat H_I, ~~~~\hat H_S=\omega_x \hat {J}_x, \nonumber \\
    & \hat H_B=\sum_{k} \omega_k \hat b_k^\dagger \hat b_k, ~~~ \hat H_I=\hat {J}_z \sum_{k} \eta_k(\hat b_k^\dagger+\hat b_k).
\end{align}
Here $\hat H_S,\hat H_B$ and $\hat H_I$ are respectively the system, bath, and interaction Hamiltonians, $\hat b^\dagger_k$ and $\hat b_k$ are the creation and annihilation operators of the $k$th mode of the bath (B), and $\eta_k$ are the coupling rates of this mode to the system,  all $\eta_k$ taken to be equal. The collective spin operators in Eq. \eqref{AtomBathHamiltonian}, $\hat {J}_i=\sum_k \hat \sigma_k^i (i=x,y,z)$, the Cartesian components of the total angular momentum $\vec{{J}}$ which is the sum of the Pauli operators of the system, $\hat \sigma_i^k$ being the Pauli operators for the $k$th TLS.  The  total Hamiltonian $\hat H$ commutes with $\hat {J}^2=\sum_i \hat {J}_i^2$, hence  the bath couples independently to each subspace of the system that has the total angular-momentum  value $j$, which ranges from 0 to $N/2$.

In order to bypass  the  intractable effect of noncommutativity of $\hat {J}_x$ and $\hat {J}_z$   in the total Hamiltonian, we switch off  the TLS level- splitting $\omega_x$, thereby eliminating  $\hat H_S$. The dynamics is then exactly solvable for any bosonic bath. For each  $j$ , the joint  evolution operator of the system and the bath is then given by
\begin{align}\label{unitaryoperator}
    \hat U_j(t)=\exp \left[ -it \Delta_{L} \hat {J}_z^2+\hat {J}_z \sum_k \left[\alpha_k(t) \hat b_k^\dagger-\alpha^*(t) \hat b_k \right]\right]
\end{align}
where the bath-dependent  functions are 
\begin{align}\label{function}
    \Delta_L(t)&=\frac{1}{t}\sum_{k}\eta_k^2 (\omega_k t-\sin \omega_k t)/\omega_k^2, \nonumber\\
    \alpha_k(t) &= \eta_k \frac{1-e^{i\omega_k t}}{\omega_k}
\end{align}
One thus arrives at  a remarkable  result: The bath-induced evolution is driven by both linear $\hat {J}_z$  and NL  $\hat {J}^2_z$ operators.  While the linear operator  gives rise to pure dephasing, the NL  operator $\hat {J}^2_z$   entangles multispin  systems.   Its effect is  a collective bath-induced Lamb shift $\Delta_L(t)$ \cite{cohen1998atom}, namely,  energy shifting  of each spin by all others via virtual-quanta exchange with the bath.

 Let us assume that each TLS is initially in a superposition of its energy $(\hat \sigma_z^k)$ eigenstates. The entire $N$-spin system then in a product of such superposition states, wherein  the individual TLS  are  {\it uncorrelated}, which has the form
\begin{align}\label{initialstate}
\rho(0) &=\ket{\varphi(0)}\bra{\varphi(0)}, \nonumber\\
    \ket{\varphi(0)}&=\ket{\theta,\phi}=\ket{\varphi_1}\otimes \ket{\varphi_2} \cdot\cdot\cdot  \ket{\varphi_N}, \nonumber\\
    \ket{\varphi_k}&=\cos\frac{\theta}{2}\ket{\uparrow}+\sin \frac{\theta}{2} e^{i\phi} \ket{\downarrow}. 
\end{align}
This  state  is an  eigenstate of the total angular-momentum/spin operator $\vec{J} \cdot \hat n$,  $\hat n$ being the unit vector on the hypersphere defined by the angles $\theta,\phi$. The decoherence rate $\Gamma(t)$ is  made negligible by dynamical control \cite{bhaktavatsala2011generation}, then the NL term $\Delta_L(t)\hat {J}^2_z$ in Eq. \eqref{unitaryoperator} evolves the initial uncorrelated state \eqref{initialstate}  analogously to  its evolution under the NL Kerr Hamiltonian \cite{yurke1986generating,Kitagawa1993PRA}, into an entangled multipartite state, which attains a non-gaussian form, becoming a macroscopic quantum superposition (MQS) state in due time.

The earliest formation time of MQS is then \cite{bhaktavatsala2011generation} 
\begin{align}
    \tau_{MQS} \equiv t=\frac{\pi}{2\Delta_L(t)}.
\end{align}
The time at which such a state forms is independent of $N$ and may be much shorter than the decoherence (dephasing) time.
This  MQS is equivalent to a  GHZ-like state in which all the $N$ spins are maximally entangled.  

 A possible realization of this effect  is by  $N$ atoms coupled to a  leaky cavity bath  \cite{schuster2008nonlinear,braun2002creation}  consisting of photonic modes.  Their  $\hat {{J}}_x$ coupling to this bath causes population exchange between the TLS  levels or their relaxation  and hampers the NL evolution. To overcome this hurdle, we have to  eliminate  the TLS level splitting, i.e. set  $\omega_z=0$ ,e.g.,  by imposing Zeeman shifts. Once the TLS  are prepared in degenerate  states $(\hat H_S=0)$,  cooperative coupling  can be induced between the cavity bath and the multiatomic system by a  Raman process and trigger the collective NL evolution  as per Eqs. \eqref{unitaryoperator},\eqref{function}.

 \begin{figure*}
\includegraphics[scale=0.4]{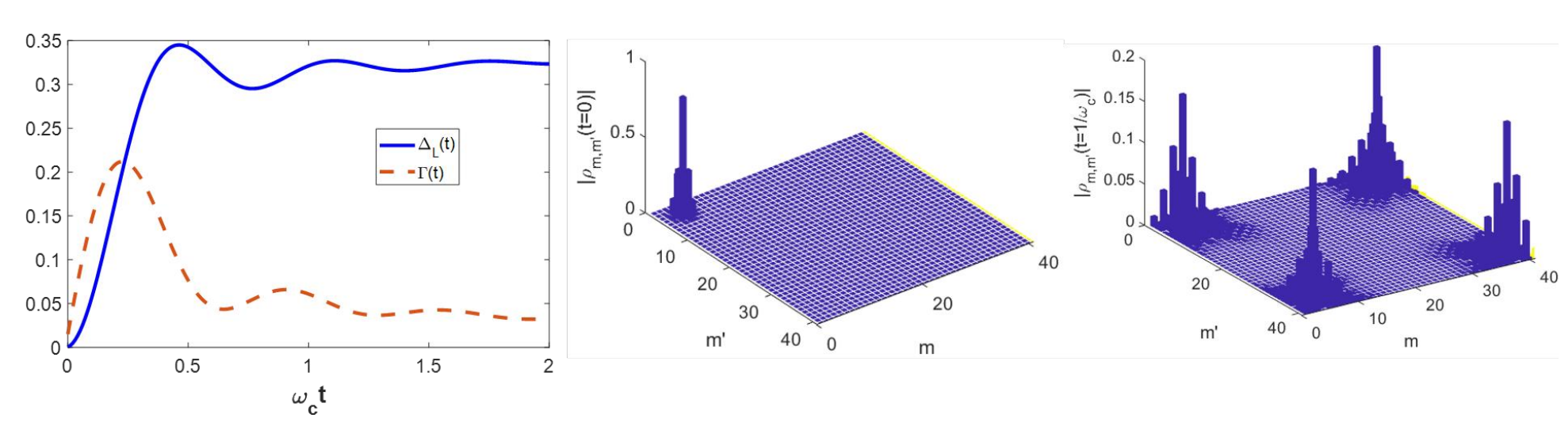}
\caption{Multiatom interaction with a cavity-field mode that has a Lorentzian coupling spectrum of width $\omega_c$ and centered at $\omega_0$. Left: $\Delta_L(t)$ and $\Gamma (t)$ are plotted as a function of time. Right: The formation of the macroscopic superposition state at times $t \sim 1/\omega_c$. For the simulations
we have chosen $\omega_c/\omega_0 = 10$ and the number of atoms $N = 20$. }\label{Durga1} 
\end{figure*}

The maximal  number  $N$ of atoms or spins  in the MQS is set by the decoherence (dephasing) that  depends on the spectral response and the temperature of the bath.  A leaky cavity acts as a Lorentzian bath whose decoherence rate  may be suppressed by dynamical control that is  optimally adapted to the bath spectrum \cite{clausen2010bath}.  Nevertheless, the residual decoherence  and its finite ratio to  the  achievable atom-cavity coupling strengths \cite{walther2006cavity,reiserer2015cavity,Meher2022EPJP} may limit the size of the  MQS allowed by condition $\tau_{MQS}\bar{\Gamma} N^2<1$ to $N <100$ (Fig. \ref{Durga1}), where $\bar\Gamma N^2 $ is the upper limit of time-averaged decay rate.  

 Finally, by extending the analysis to 3-level (Lambda) atoms in the cavity, we can map the multiatom  MQS to a corresponding entangled state of a two-mode field described by Stokes operators: Under electromagnetically-induced transparency (EIT) conditions  in the cavity, the mapping is described by the effective Hamiltonian \cite{dantan2004quantum} 
 \begin{align}
     \hat H=g_{eff}\vec{{J}}\cdot \vec{S}.
 \end{align}
Here, $\vec{{{J}}}$ and $\vec S$ are, respectively, the multiatom angular-momentum and the photonic  two-mode Stokes operator, effectively coupled with strength $g_{eff}$.

The  exactly solvable model  surveyed above  reveals  the unexpected entangling dynamics of a multiatom or multispin system by a  thermal bosonic  bath to which it is linearly, collectively coupled. This intriguing consequence of  a commonly occurring interaction of a quantum system with a finite-temperature environment may  induce  the deterministic  formation of entangled, non-gaussian, possibly MQS states of the system.These states may then be mapped onto a two-mode photonic  MQS   of a field incident on the cavity. 

If the incident field is thermal, the MQS will be averaged over the thermal distribution in the coherent-state basis. This would hamper but not necessarily destroy the MQS, provided the thermal distribution width is smaller than the  decoherence rate of the cavity bath. 

It follows from the above discussion that the multiatom interaction with the cavity bath can become  an \textit{entangling} resource for initially uncorrelated (unentangled) atoms and (subsequently) two-mode fields. Despite the technical complexity, this approach appears to be a viable route to NL quantum thermodynamics.

\section{From deterministic to probabilistic NL transformations}  
\label{Sec5}
Thus far, we have described deterministic, autonomous NL transformations or filtering of  thermal noise to non-gaussian states  with advantageous  properties: ergotropy  for heat engines  or information for quantum  sensing   and microscopy. The open question is: what practical schemes can be made available for the implementation of such NL transformations in the few-photon domain beyond the  technically demanding (albeit feasible) schemes discussed in Sec. \ref{Sec4} ?  

Alternatively, NL-transformed states can be obtained via the system measurements by a quantum probe, which is coupled to the system by a simple interaction Hamiltonian. Post-selected events, known as conditional measurements (CMs), can prepare the targeted states with limited success probability, which can be optimized.  

 In the quest for promising measurement-based schemes for NL thermal state transformations,  we may be motivated by  the recent experimental demonstration  that thermal state of a spin bath can be filtered into nearly-pure states by a series of CMs \cite{dasari2022anti} performed on a probe spin coupled to the spin bath. 
 The time-intervals between probe measurements chosen such that the excitation swap of the spin bath and the probe was maximal, conforming  to the anti-Zeno (AZE) regime \cite{kurizki2022Book,kofman2000acceleration}. Only those measurement outcomes were post-selected wherein the swap occurred, to which only spins with certain frequencies contributed. We thus collapsed/filtered the thermal-spin bath state to a nearly pure collective state.  Another measurement-based scheme proposed by our group has been recently demonstrated as a probe of single-photon polarization- noise correlations \cite{virzi2022quantum}. 
 
 Can one realize  analogous methods to achieve  NL filtering of multiphoton noise? In what follows, we highlight several prospective measurement schemes for  photonic thermal-state filtering aimed at realizing non-gaussian photon states, particularly Fock states.

\subsection{NL transformations of a field mode via photodetection:}
\begin{figure}
\includegraphics[scale=0.55]{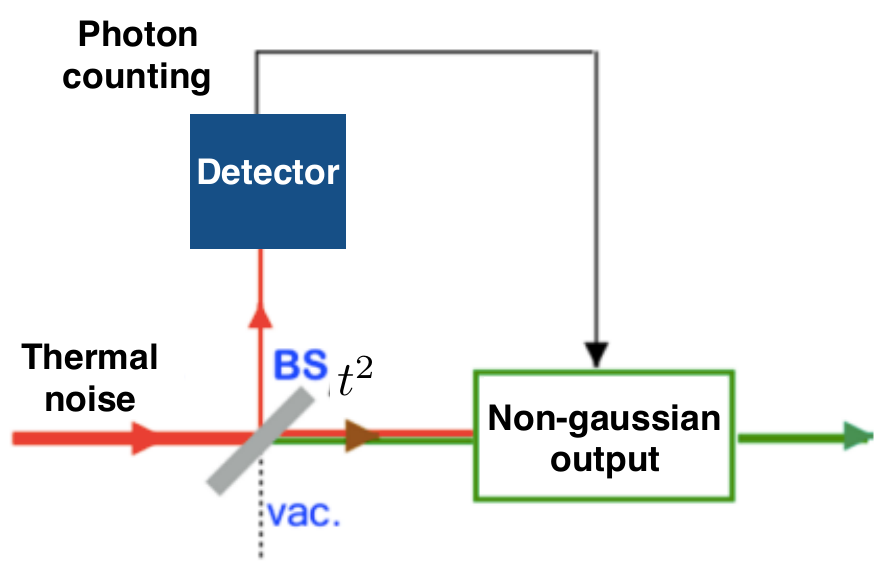}
\caption{NL transformation of thermal input by photocount on a small fraction of a noisy input signal reflected by a BS with high transmissivity $t^2$.}\label{photocoun-scheme} 
\end{figure}
NL  transformation of a field mode in a thermal state can be achieved by photon counting on a small fraction of the input field, which constitutes a CM of the field state in the Fock basis \cite{mePRE}. The measured small fraction is the reflected part  (Fig. \ref{photocoun-scheme}) of the input beam  incident on a BS with high transmissivity $t^2$.

\begin{figure}
\includegraphics[scale=0.4]{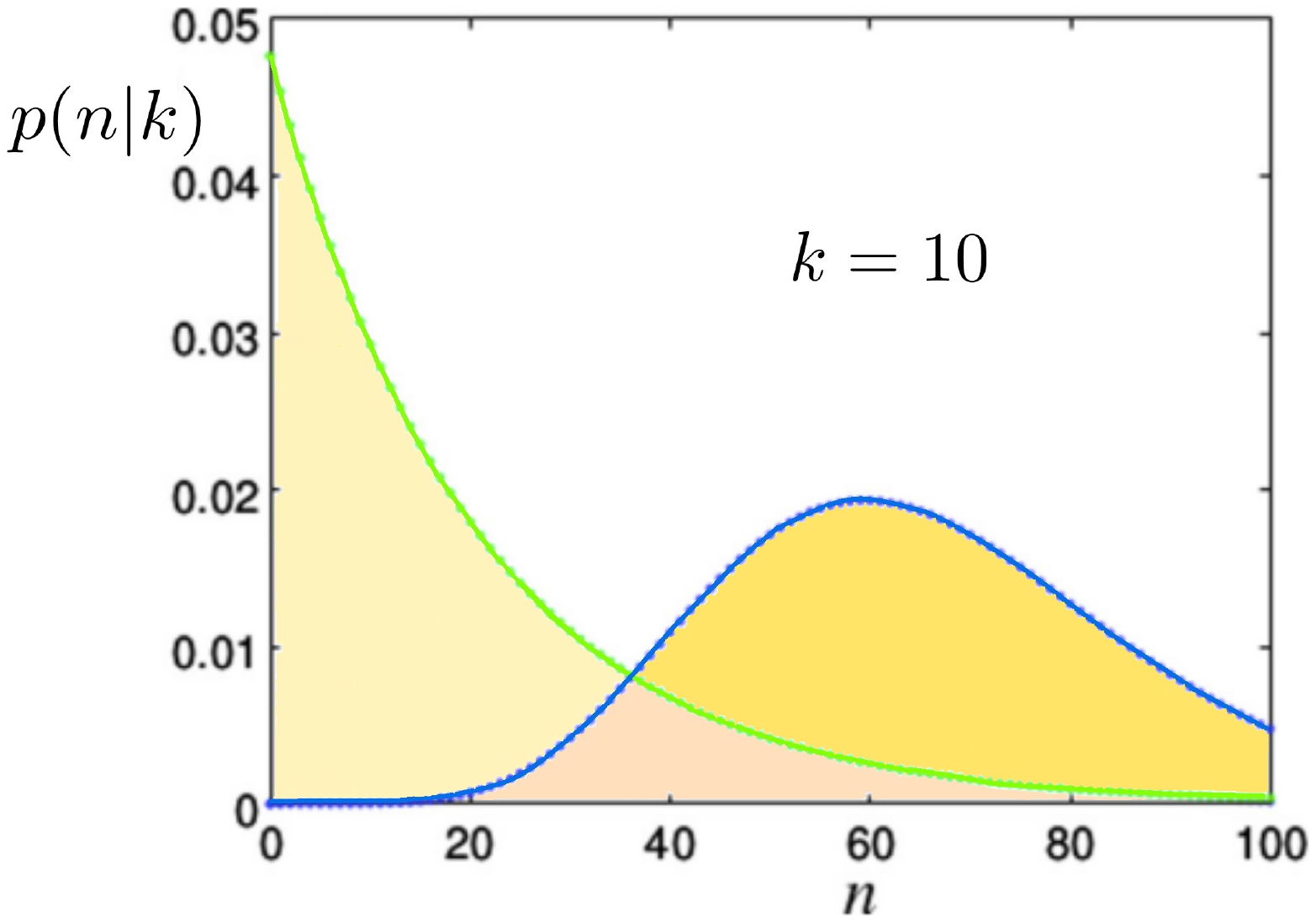}
\caption{Fock state distribution in the postmeasured output state for $k=10$ photon detection. The input thermal state has mean number of photon $\bar n=20$. BS has transmissivity $t^2=0.9$. The green and the blue dotted lines represent the initial thermal and postmeasured Fock state distribution respectively.}\label{Fock-photocount} 
\end{figure}
\begin{figure}
\includegraphics[scale=0.5]{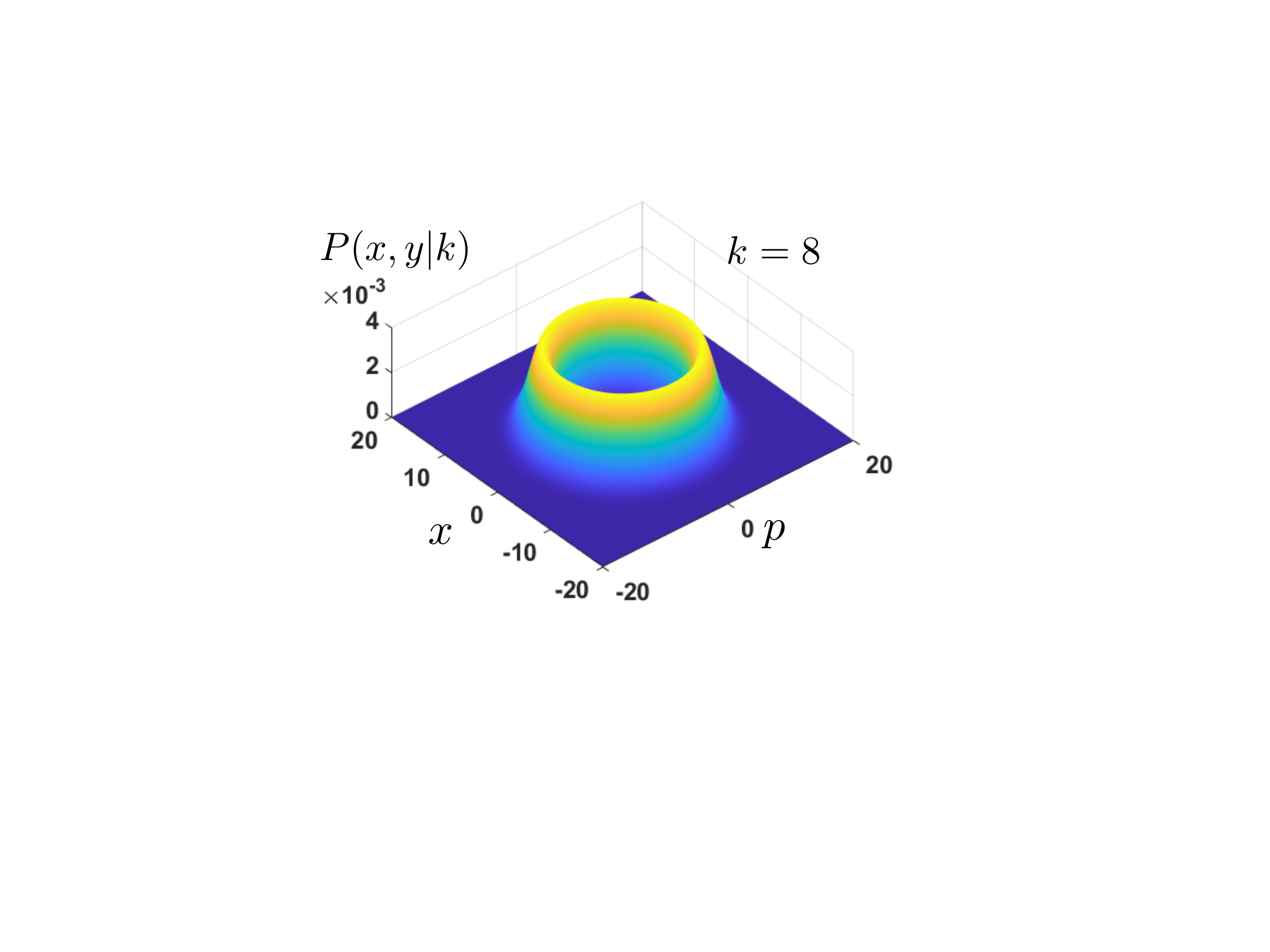}
\caption{ The  phase-plane probability distribution of the output state conditioned on measuring $k=10$ photons for a thermal input state having $\bar n=20$ and BS transmissivity $t^2=0.9$. The nonmonotonic character of the phase-space output distribution corresponds to a non-gaussian, non-passive state.}\label{photocount} 
\end{figure}
The detection of $k$ photons in the reflected fraction  yields the  measurement-transformed field state which is transmitted through the BS 
\begin{eqnarray}
\label{fockpm}
 \rho_{k}&=\sum_{n=0}^\infty P(n|k)|n\rangle \langle n|.
\end{eqnarray}
The  non-thermal measurement-transformed output state depends  on $\bar n$, $t^2$ and the CM outcome $k$ (Fig. \ref{Fock-photocount}). The projection of the reflected beam on the $k-$photon state is described in the $x$-$p$ plane by a ring that cuts out a hollow from the gaussian distribution of the input state (Fig. \ref{photocount}). The nonmonotonic character  of this ring-shaped outcome state is an evidence of its nonpassivity.  Namely, the state in Eq. (\ref{fockpm}) is passive, by definition, iff the the Fock state populations obey the monotonic falloff as a function of $n$ \cite{Allahverdyan2004EPL}
\begin{equation}
\label{passive-condi}
 P(n|k)\geq P(n'|k),~~~\forall n,n',
\end{equation}
but this condition is violated for $k\neq0$ measurement results.
The corresponding output state is found to be nonthermal upon examining the second-order coherence (autocorrelation) function \cite{Carmichael_BOOK,Gardiner,ScullyZubairy} of the state in Eq. (\ref{fockpm}), which has the form
\begin{eqnarray}
\label{g2}
 g^{(2)}(0)
 &=& \frac{\sum_n P(n|k)~n (n-1)}{\bar n_k^2}\\ \nonumber
 &=& 1+ \frac{1}{1+k}.
\end{eqnarray}
This expression does not depend on the thermal  input  temperature or the  BS splitting ratio. Only when $k=0$  photons  are counted,  the entire beam is transmitted  through the BS in a thermal state, with $g^{(2)}(0) = 2$. In the opposite  limit of many detected photons, $k >> 1$, Eq. \ref{g2} approaches  $g^{(2)}(0) = 1$, corresponding to a Poissonian distribution.

The average energy of the post-measured output state,
\begin{equation}
  E_k= \hbar \omega \frac{(1+k)t^2}{e^{\frac{\hbar \omega }{k_B T}} -t^2}
 \end{equation}
can yield the following ergotropy (work capacity) after subtracting the passive part of the output,
 \begin{equation}
  W_k=E_k-E'_k.
 \end{equation}
Averaging over all possible measurement results  $k$ yields the extractable mean work output
 \begin{equation}
  \bar W=\sum_k P_k W_k,
 \end{equation}
which signifies the non-passivity of the overall CM-transformation.
\subsection{NL transformations of a field mode via homodyne measurements}
\begin{figure}
\includegraphics[scale=0.45]{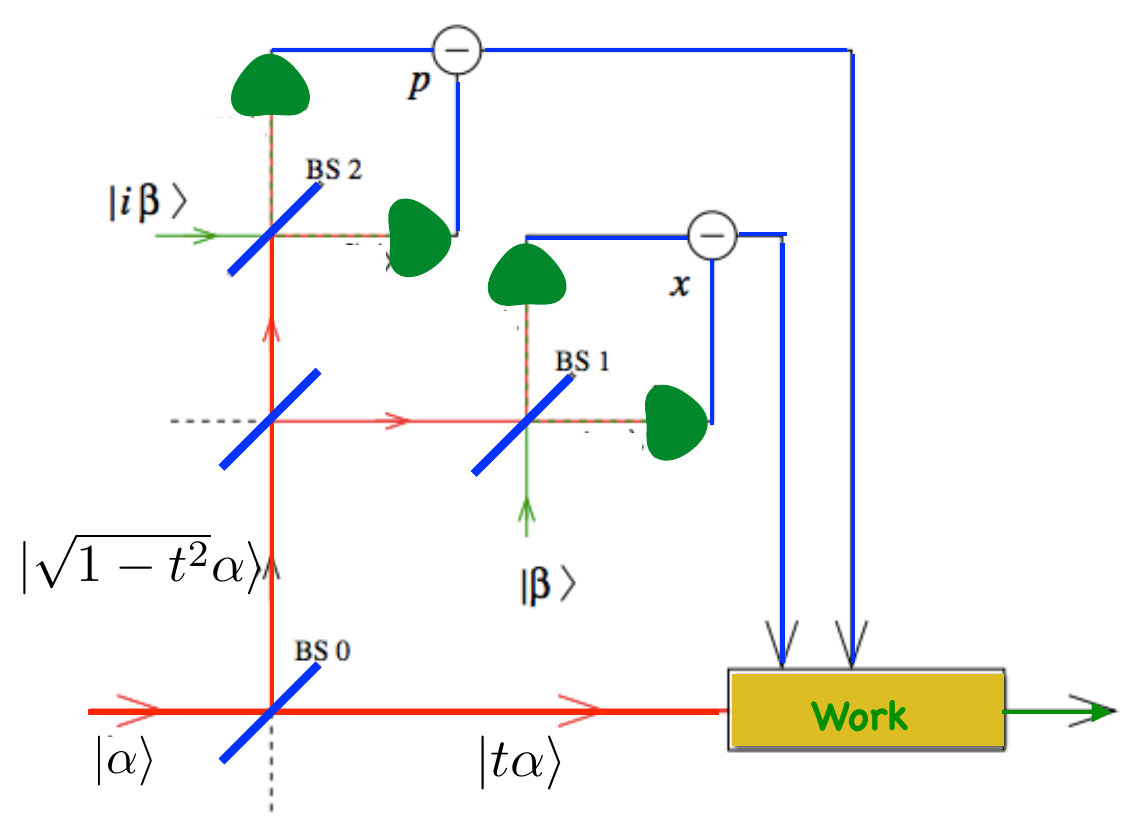}
\caption{ Homodyne measurement of a small fraction (reflected at BS0) of input state $|\alpha\rangle$. It is mixed with LO quadratures $|\beta\rangle$ and $|i\beta\rangle$ at homodyne setups that measure $\Delta n_x$ and $\Delta n_p$. The postmeasured output (transmitted at BS0) is generally non-passive and can deliver work.}\label{homodyne} 
\end{figure}

Homodyne phase-sensitive measurement of a small reflected fraction constitutes a highly effective CM transformation of a thermal input state \cite{OpatrnyPRL21}. 
Let us first consider in Fig. \ref{homodyne}  a  weak  input signal in a coherent state $|\alpha\rangle$ with  amplitude $\alpha = \frac{1}{\sqrt{2}}(x + ip)$. It  is weakly  reflected by a BS towards a homodyne-measurement setup  where its quadratures are combined with those of local oscillators (LO) with real coherent quadrature-amplitude $\beta$ and imaginary quadrature amplitude $i\beta$. The two  homodyne detectors that measure  the orthogonal quadratures of the signal combined with the LO yield     photocount differences $n_x$ and $n_p$  which  provide information on the input-field quadratures $x$ and $p$ \cite{Opatrny2015PRA2}. 

\begin{figure}
\includegraphics[scale=0.5]{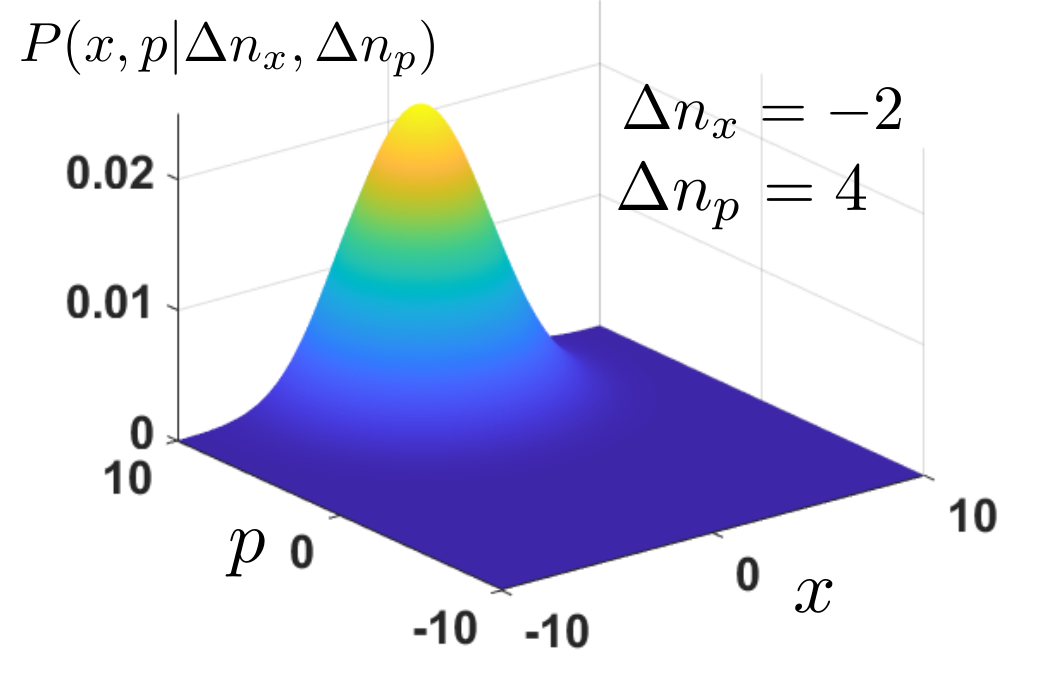}
\caption{ The phase-space probability distribution of an homodyne measured output state conditioned on homodyne measurement results $\Delta n_x$ and $\Delta n_p$  for thermal input state having $\bar n=25$, BS transmissivity $t^2=0.8$, and LO mean amplitude $\beta=1.6$. }\label{homodynephasespace} 
\end{figure}

The same setup can be used to transform  a  thermal  input state with a mean number of photons  $\bar{n}$,  which can be represented as \textit{a random mixture of coherent states} $\vert \alpha \rangle$   with  a gaussian P-distribution. The  post-measured distribution of $\alpha$, $P(\alpha |\Delta n_{x},\Delta n_{p}) $  is conditioned on the detection of photon-number differences $\Delta n_x$ and $\Delta n_p$ 
yielding the output state  (Fig. \ref{homodynephasespace})
\begin{eqnarray}
\hat \varrho (\Delta n_{x},\Delta n_{p}) 
&=& \frac{1}{t^2} \int \! \int P\left(\frac{\alpha}{t}|\Delta n_{x},\Delta n_{p}\right) |\alpha \rangle \langle \alpha | d^2 \alpha \nonumber \\
\label{varrhoDn}
\end{eqnarray}
The mean work output is obtainable upon averaging the ergotropy over all possible $\Delta n_{x},~~\Delta n_{p}$  and subtracting the energy of the two LOs, $2 \hbar \omega\beta^2$. The expression can be optimized with respect to the LO amplitude and  the BS transmissivity resulting in the maximal work extractable from the post-measured output is 
for a thermal state with large $\bar n$, the optimized BS transmisivity and LO energy are $1-t^2= \frac{1}{\sqrt{\bar n}}$ and $2 \beta^2= \sqrt{\bar n}$, respectively, yielding the maximal extractable work (in $\hbar \omega$ units) \cite{OpatrnyPRL21}
\begin{eqnarray}
W_{\rm max} \approx
 \bar n- 4\sqrt{\bar n}+ 6.
\label{Woptimum-large-n}
\end{eqnarray}

  The information gain in this small-fraction homodyne transformation is characterized by the mutual information \cite{SagawaPRL08,Vid2016} that has to be processed for work extraction via CM feedforward.  For maximum work extraction as in Eq. \eqref{Woptimum-large-n}, we obtain in the limit  $n \gg 1$ 
the total mean mutual information gain 
\begin{eqnarray}
\label{IH}
\mathcal{I}\approx \frac{1}{2} \ln \frac{\bar n}{4}. 
\end{eqnarray}

Hence, the cost of feedforward required for work extraction via CM based on the information gained (Eq. \ref{IH}) becomes negligible compared to output (Eq. \ref{Woptimum-large-n})  for large $\bar n$. This is  the main advantage of work extraction obtained by measuring a small fraction of the input, either by photocount or by homodyning (compare with \cite{Vid2016}). In any case, the P-distribution of the post-measured state is non-thermal, unless both $n_x$ and $n_p$ are measured to be zero. These measured results control the character of the output state.

  Finally, if the results of the  measurements  discussed above are unread, namely, we resort to nonselective measurements (NSMs) \cite{kurizki2022Book,NielsenBOOK00}, the output  state can be shown to remain passive and  yield no work, neither  via  small-fraction homodyne detection nor via small-fraction photodetection \cite{mePRE}. Nevertheless, NSMs that are executed by operators that are capable of extracting work \cite{niedenzu18quantum,Talkner1}.
  
  Remarkably, when the modes are nonlinearly correlated, NSMs can yield work from the intermode correlation energy \cite{Erez2008,GelbwaserPRA13}, an effect  that goes beyond  Landauer's principle \cite{LandauerIBM61}.
  
  Thus, measurements can transform thermal input states into an ergotropy/work-capacity resource.

  \subsection{NL transformation of thermal cavity fields by measurements on resonant atoms}
Early on,  we showed that CMs  of  the atomic state can yield,  on demand,  diverse states of a cavity-field mode that is  coupled to the atoms, if this field mode initialized in a thermal \cite{harel1996fock} or coherent state \cite{kozhekin1996quantum, harel1996optimized, Garraway1994PRA}. This procedure could be realized by multiple two-level  atoms transmitted through an ultrahigh-Q cavity, where they interact with the single-mode field via the resonant Jaynes?Cummings Hamiltonian \cite{ScullyZubairy,cohen1998atom,walther2006cavity,Meher2022EPJP} or its off-resonant (dispersive) \cite{Gerry} variant.  Control of the output cavity field state was shown to be enabled in these schemes by  adjusting the time-intervals between consecutive atoms in the cavity.  This measurement strategy can transform thermal input photon distributions into any chosen  Fock state with up to a few tens of photons. Such transformation requires in this strategy can be accomplished in a time that is much shorter than the lifetime of the prepared Fock state.  

The strategy is based on intermittent, alternating,  application of two types of  manipulations:

\noindent
\textbf{1. Nonselective measurements (NSMs):} We  excite the atom and send it through the cavity at  a fixed velocity. The atom-field interaction, followed by a NSM on the atom, which amounts to  disregarding the measured outcome, following transformation of the number (Fock-) state populations in the cavity
\begin{equation}\label{PopilationInCavity}
    \rho_{nn}\rightarrow \rho_{nn}\cos^2(g \tau \sqrt{n+1})+\rho_{n-1,n-1}\sin^2(g \tau \sqrt{n}),
\end{equation}
$\tau$ being the cavity traversal time, and $g$  the field-atom coupling constant. 

 According to Eq. \eqref{PopilationInCavity}, each number state losses population to the number state that has one more photon via stimulated emission from the atom with probability $\sin^2 (g \tau \sqrt{n+1})$. The outcome is a nonuniform drift of the  Fock-state distribution  towards increasing  $n$. This drift is minimal for $n$ that  satisfy 
\begin{equation}
\label{integer}
  g \tau \sqrt{n+1}\approx m \pi  
\end{equation}
for an  integer $m$. When $g\tau$ and $m$ are chosen to satisfy Eq. (\ref{integer}), the population of the corresponding $\vert n \rangle$  may grow  at the expense of number states having smaller $n$, since  Eq. \eqref{integer} preclude photon emission.  This $\vert n \rangle$ may be deemed a "trapping" state.

\begin{figure*}
\includegraphics[scale=0.4]{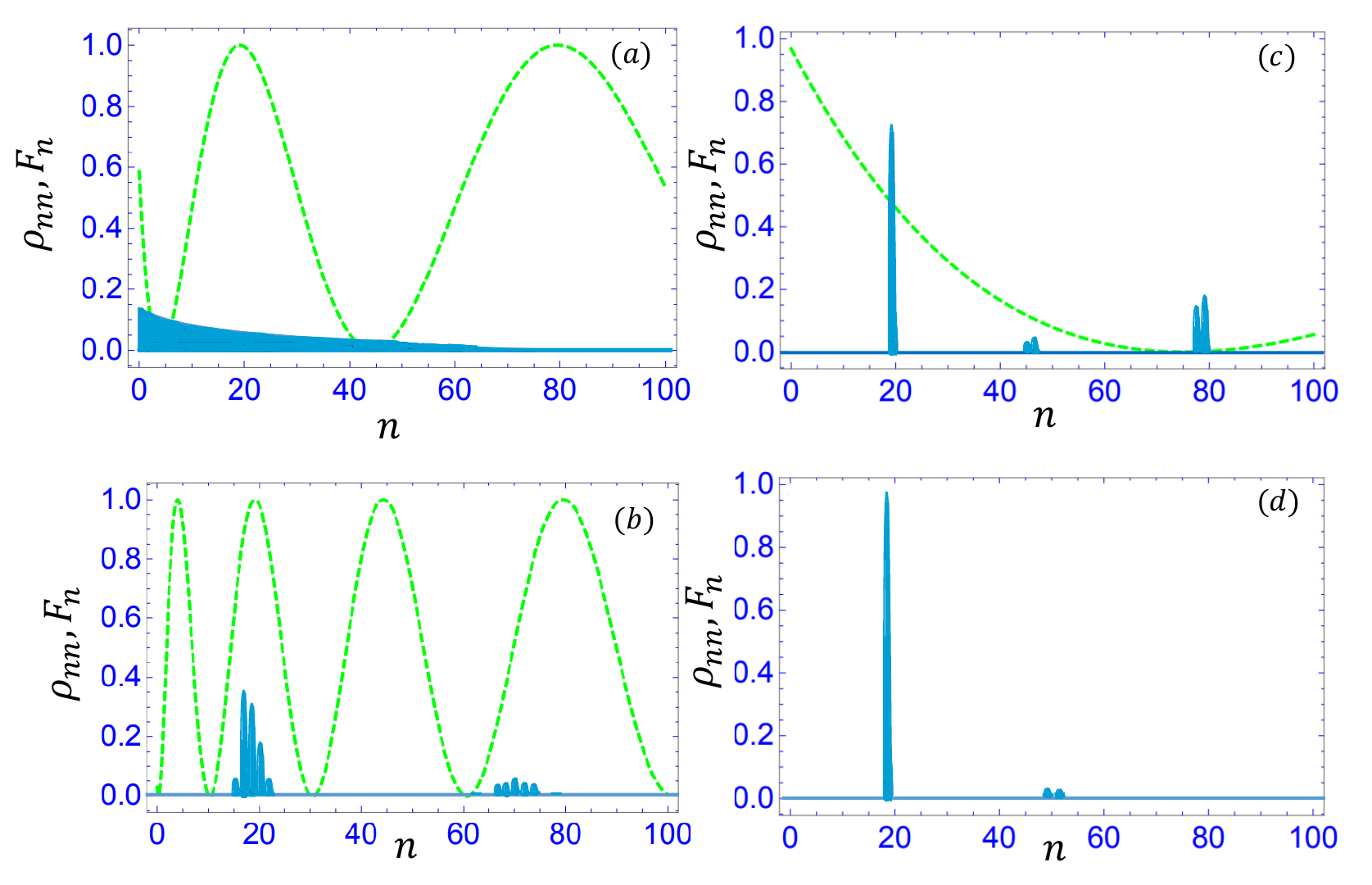}
\caption{Evolution  of an initial thermal photon-number distribution $\rho_{nn}$ in a cavity towards a target Fock state $n$ via non-selective and conditional measurements (NSM and CM) of initially excited atoms governed by the resonant Jayness-Cummings interaction Hamiltonian. Dashed green: NSM filter function $\cos^2(g\tau\sqrt{n+1})$ where $g$ and $\tau$ are the field-atom coupling strength and interaction time, respectively. Solid blue: $\rho_{nn}$ distribution; (a) Initial $\rho_{nn}$ (b) $\rho_{nn}$ after $100$ NSMs for $\tau_1$ chosen by Eq. \eqref{integer} (c) $\rho_{nn}$, after $50$ NSMs for $\tau_2=2\tau_1$. (d) Same, after one CM with $\tau_3$ that satisfies Eq. (28)}\label{Harel1} 
\end{figure*}

The populations of $\ket{n}$ in the range between consecutive states that satisfy the "trapping" condition  are shifted as a results of an NSM that has the larger  $n$. By iterating  this shift, via multiple consecutive NSMs,  one gradually concentrates the Fock-state  distribution near the trapping states {(Fig. \ref{Harel1})}.
Yet, this  concentration process is progressively slowed down, since the  efficiency of shifting the distribution decreases  for $\vert n\rangle$  near  the ith trapping state $\vert n_i\rangle$. We can however  accelerate the concentration upon reducing the atomic velocity by one half, thus doubling the traversal time $\tau$ and $m$ in Eq. \ref{integer}. The  concentration process then  regains its efficiency while keeping the velocity selection simple {(Fig. \ref{Harel1})}. 

\noindent
\textbf{2. Conditional measurements (CMs):} NSMs alone  cannot convert a  thermal number-state distribution into a single number state, since  they accumulate the number-state populations  in the vicinity of several $n$ values thatt satisfy the trapping condition. As a remedy, we employ a \textquotedblleft filter" based on a CM that erases portions of the distribution resort to near trapping states populated by the NSMs with the exception of number states we target. To this end, we send an atom in the excited state and measure whether, upon exiting  the cavity, it is still in the excited state. The success probability is given by 
\begin{equation}
    \label{prob}
    P_e=\sum_{n=0}^\infty \rho_{nn}\cos^2(g \tau \sqrt{n+1}).
\end{equation}
In this case the Fock-state populations are transformed as
\begin{equation}
    \rho_{nn}\rightarrow P_e^{-1}\rho_{nn}\cos^2(g \tau \sqrt{n+1}).
\end{equation}
If the CM fails, namely, we should resend another excited atom, and measure its state after it exits the cavity. Instead of Eq. (\ref{integer}), this CM requires 
\begin{equation}
\label{halfinteger}
  g \tau \sqrt{n+1}=(m +\frac{1}{2})\pi  
\end{equation}
so that  the CM suppresses  (\textquotedblleft erases\textquotedblright !) the populations of Fock states near $\vert n \rangle$  for a chosen $n$ and some integer $m$ (Fig. \ref{Harel1}). 

A plausible   experimental scenario starts from  an initial thermal distribution of number states in the cavity, corresponding to the finite temperature of the cavity. The efficiency with which the field is transformed from such a  broad  thermal distribution to a very   narrow distribution near the target Fock state is reflected by  the Shannon entropy of the field, 
\begin{equation}
    \label{entropy}
    S=-\sum_{n=0}^\infty \rho_{nn}\log \rho_{nn}
\end{equation}

\begin{figure}
\includegraphics[scale=0.45]{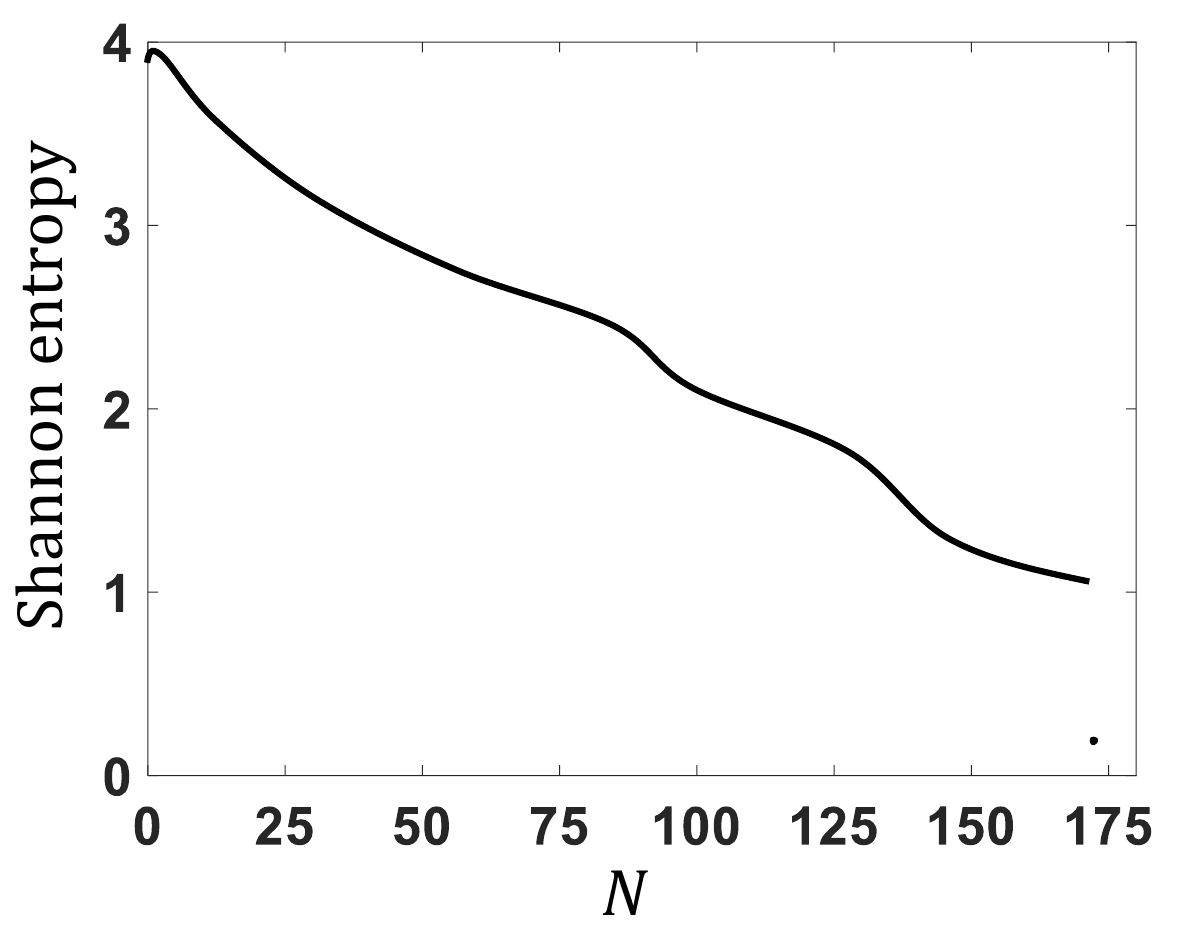}
\caption{Decreasing Shannon entropy of the output state in the cavity as a function of NSM numbers $N$ in all  the sequences completed by a successful CM.}\label{Shannon} 
\end{figure}

The Shanon entropy evolution  allows us to monitor the entire preparation process  (Fig. \ref{Shannon}). It shows that by combining nonselective and conditional measurements (NSMs and CMs) on sequences of excited atoms sent through  cavity, thermal  high-entropy state can be transformed into a targeted number state with high probability after few measurements. 

\section{Discussion} \label{Sec6}
We have surveyed the  achievements and the potential  of the recently introduced nonlinear (NL) thermodynamic effects  in coherent, nondissipative systems, comparing the quantum and classical aspects of  these effects. We have mostly dwelled on its conceivable implementation schemes and  their technological prospects.
The exploitation of NL intermode correlations for filtering or transforming thermal noise input into non-thermal  output with controllable characteristics opens new perspectives for the realization of diverse thermodynamic devices with novel functionalities. 

Apart from unconventional NL  coherent heat engines \cite{Opatrny2023ScAdv}, quantum  microscopes \cite{meher2024supersensitive} and  quantum noise sensors \cite{meher2024thermodynamic} discussed in Sec. \ref{Sec2}, other promising applications include NL  quantum  thermometers \cite{mukherjee2019enhanced}, as well as heat diodes and heat  transistors \cite{PhysRevE.99.042121,Tahir,segal2005spin,saira2007heat,segal2008single,shen2011single} that may benefit from NL filtering of the thermal input.  

A more ambitious, but not impossible, undertaking would be the construction  of  multi-node networks comprised of NL-coupled elements: such networks would be capable of performing classical calculations even with thermal input \cite{reiserer2015cavity,miller2024nonlinear}.
 Operation   of   NL coherent  thermodynamic  devices  would not be prohibitively hard in the quantum domain, provided the goal is only to keep track of heat and ergotropy exchange between the modes, which are compatible with coarse graining  of the evolution and are thus  much less challenging than quantum computations that rely on multi-element entanglement \cite{NielsenBOOK00} (Sec. \ref{Sec3}). As we have shown, the salient  quantum feature of thermal-noise NL filtering is the ability to control the output  photon statistics  and render the output state  non-gaussian. 
The bottleneck impeding the realization of such NL thermodynamic devices is the need for giant nonlinearities that may yield large phase shifts of each  evolving field mode by cross coupling it with only few quanta in other modes. At present, conversion of photons into dipole-dipole interacting  Rydberg polaritons in a cold gas \cite{Friedler,drori2023quantum}  (Sec. \ref{Sec4}) is the only demonstration of giant NL cross-Kerr effects between field modes populated by one or  few  photons.  Dispersive field-atom interaction in an ultrahigh-Q cavity \cite{walther2006cavity,reiserer2015cavity,Meher2022EPJP}  can also exhibit  a cross-Kerr effect with single photon per mode but cannot be readily incorporated in practical devices. 

Yet,  other  promising  technologies  may be conceived  with the goal of  realizing  few-photon NL devices: 
\begin{itemize}
    \item 	Multiatom interaction with fields in a moderately high-Q cavity: Since such a  leaky cavity  constitutes a photonic bath, its  dispersive (off-resonant) interaction with multiple atoms is a realization of a scenario in which NL unitary  effects, combined with dynamical control \cite{clausen2010bath},  transform an $N$-atom product state into a Schroedinger-cat state, alias GHZ or MQS state \cite{bhaktavatsala2011generation}. Subsequently, such a state can act as a massive entanglement resource  that can transform   two independent field modes  that interact with the cavity  into an entangled two-mode N-photon state. (Sec.4). The  feasibility of high-fidelity  MQS entangled states with $N\leq 100$ may promote quantum sensing as well as one-way quantum computing \cite{briegel2009measurement}.

    \item   One may revisit the proposal for giant Kerr cross-coupling via interaction of  a self-induced transparency (SIT) soliton in a resonantly-doped Bragg structure with a pulse subject to electromagnetically-induced transparency (EIT) \cite{kurizki2002nonlinear}. Such an SIT soliton can have an arbitrarily small pulse area and slow velocity that matches the EIT-pulse velocity, so that the two pulses interact over long distances.

    \item  As a substitute for the above deterministic, coherent approaches,  one may employ probabilistic measurement-based state preparation that can emulate NL filtering of thermal noise. Such methods may include  small-fraction photodetection (Sec. \ref{Sec5}A)  or homodyning measurements (Sec. \ref{Sec5}B), which allow for thermal-state  transformation into  a non-Gaussian state conditioned on the measurement result, or the interaction of a thermal state in a cavity with a sequence of atoms which combines non-selective and conditional measurements, resulting in an output state that resembles a Fock state (Sec. \ref{Sec5}C). Although the main goal of NL thermodynamics is to be the basis for coherent device operation, such \textquotedblleft poor man's\textquotedblright substitutes  may have  instructive merit  and achieve similar device functionalities, at the price of dependence on the success or failure of the underlying measurements. 

    \item  As a curiosity, we note that the interaction of an atom with an empty cavity freely falling into a black hole can yield thermal  photonic states, as if the cavity were immersed in a thermal bath, although both outside and inside the cavity there is complete vacuum \cite{misra2024black}. The reason for such a paradoxical effect is the nonlinear transformation of the vacuum in the vicinity of a black hole. This transformation may also be classified as an exotic kind of NL thermodynamics.
    \end{itemize}
To conclude, we hope that the nascent, unconventional  theme of NL thermodynamics, which poses a  broad range  of  intriguing  conceptual and practical questions, will be further developed, both theoretically and experimentally. 
To this end, a variety of platforms and approaches may be harnessed, with a view of replacing  the established paradigms of thermodynamic device composition and functionality by a new approach that emphasizes coherent, autonomous operation of such devices.

\vspace{1cm}
 
\noindent
\textbf{Data Availability Statement:} No data associated in the manuscript.

\vspace{0.5cm}
\noindent
\textbf{Acknowledgements:} G.K. and D.B.R.D. acknowledge support from the DFG under grant number FOR2724. A.M. acknowledges support from the Anusandhan National Research Foundation, Government of India, under grant number ANRF/ECRG/2024/003836/PMS.

\bibliography{QNLT}

\end{document}